\documentclass[pra,showpacs,preprintnumbers,amsmath,aps]{revtex4}
\usepackage{graphics}

\def\be{\begin{equation}}
\def\ee{\end{equation}}
\def\bea{\begin{eqnarray}}
\def\eea{\end{eqnarray}}
\def\bse{\begin{subequations}}
\def\ese{\end{subequations}}
\def\bma{\begin{mathletters}}
\def\ema{\end{mathletters}}
\def\C{\hbox{$\mit I$\kern-.6em$\mit C$}}

\begin{document}

\title{Universal Quantum Filter}
\author{Dong Yang}%
\email{dyang@ustc.edu.cn}
\author{Sixia Yu}%
\email{yusixia@ustc.edu.cn}

\affiliation{Department of Modern Physics, University of Science and Technology of China, Hefei, Anhui 230026, People's Republic of China}

\date{\today}

\begin{abstract}
Universal quantum filter (UQF) is introduced and proved to exist. Optical realization of UQF is proposed in experiment.
\end{abstract}


\maketitle

\section{Intrudoction}
Quantum entanglement is responsible for many quantum tasks and most known implementations of quantum protocols require maximally entangled state (MES) \cite{NC}. Unfortunately, due to interaction with environment, entanglement is fragile and easy to be blurred by the noise, which results in non-MES pure state or entangled mixed state. To obtain MES, entanglement concentration or distillation is required \cite{Bennett2}. Often conversion between entangled states is manipulated by local operation and classical communication (LOCC). Among local operations, the filtering method is applied frequently, sometimes even essentially. Hidden quantum non-locality of some mixed states is revealed by local filter \cite{Popescu,Gisin}, which they do not violate any Bell-CHSH inequality while they do after proper local filters. The well-known entanglement distillation protocol---BBPSSW \cite{Bennett2} can not directly applied to all two-qubit inseparable mixed states. Indeed inseparable states with fidelity $f=\max_{\phi}\langle\phi|\rho|\phi\rangle\leq 1/2$ can not distilled by BBPSSW protocol where $\phi$ is a MES. However it is demonstrated that any inseparable two-qubit state can be distilled to a singlet form \cite{Horodecki}. To apply BBPSSW protocol, filtering process is performed to promote the fidelity beforehand \cite{Horodecki}. Also local filtering is an essential step to optimally concentrate entanglement or maximize the fidelity of teleportation of the mixed state \cite{Vaetraete}. A two-dimensional system---qubit is the elementary system in quantum information theory. Conversion between two-qubit entangled states are studied with detail due to its simplicity and importance. Our discussion of quantum filter is limited to qubit.

A qubit filter is a kind of two-outcome POVM. The realization of the POVM requires an interaction unitary evolution between an ancilla and the system. However, it is hard to perform arbitrary interaction unitary gate between two qubits. The interaction gate is a kind of expensive resource. So an instant question arises: {\it does there exist an interaction unitary gate such that any filtering process can be performed by modulating the valuable interaction gate with cheaper qubit unitary gate and projective measurement? } Here we assume that arbitrary qubit unitary gate and projective measurement is freely available---actually it is much easier to perform these operations. Such an interaction gate is defined as a universal quantum filter (UQF). If it exists, we can use the UQF to carry out any filtering task. Only one standard device is needed in the toolbox. If it does not, we have to construct a specific filter for a specific filtering task so that our toolbox would be full of all types of filters. In this paper, we will solve this problem. The good news is that the answer is YES. Fortunately enough, a simple choice is the CNOT gate.

\section{universal quantum filter}
A qubit filter is a POVM described by $\{N_{0},N_{1}\}$,
\be
\label{filter}
N_{0}=V_{0}\left[\begin{array}{cc}\cos\alpha & 0\\0 &1\end{array}\right]U,~~~ N_{1}=V_{1}\left[\begin{array}{cc}\sin\alpha & 0\\0 &0\end{array}\right]U,
\ee
where $V_{0},V_{1},U\in SU(2)$ and the completeness relation $N_{0}^{\dagger}N_{0}+N_{1}^{\dagger}N_{1}$=1 is satisfied. Explicitly, the same $U$ can be dropped since arbitrary qubit unitary operator is allowed before the filtering process. In the same way, $V_{0},V_{1}$ can be also neglected since arbitrary unitary transformation can be performed dependently on the measuring outcome. So a qubit filter is briefly described as $F(\alpha)$ by the two diagonal matrices. A universal qubit filter (UQF) is an interaction two-qubit unitary gate $U_{EA}$ such that any filter $F(\alpha)$ could be realized by $U_{EA}\circ (U_{E}\otimes U_{A})$ supplemented with proper projective measurement on the ancilla qubit $E$. In the following, we clarify the main idea by the optimal conversion between pure entangled states. Since the local filtering method is very useful in the entanglement distillation \cite{Horodecki,Vaetraete}, UQF can also been applied to mixed entangled states.

Suppose two parties, traditionally called Alice and Bob, initially share an entanglement state $|\phi\rangle^{AB}=\alpha_{0}|00\rangle+\alpha_{1}|11\rangle$, where $\alpha_{0}\ge\alpha_{1}\ge 0$, $\alpha_{0}^{2}+\alpha_{1}^{2}=1$. They want to transform to a more entangled state by LOCC, $|\psi\rangle^{AB}=\beta_{0}|00\rangle+\beta_{1}|11\rangle$
and $\beta_{0}\ge\beta_{1}\ge 0$, $\beta_{0}^{2}+\beta_{1}^{2}=1$. Usually the goal state is a MES that is the requirement in most quantum protocols. Of course the successful transformation is probabilistic since the average entanglement is non-increasing under LOCC. The optimal conversion means that the greatest probability of success is achieved, which is asserted by the theorem in \cite{Vidal}. In the simple case of two-qubit, it is expressed as: if $\{\alpha_{0}^{2},\alpha_{1}^{2}\}\succ\{\beta_{0}^{2},\beta_{1}^{2}\}$, $p_{max}(\phi\to\psi)=\alpha_{1}^{2}/\beta_{1}^{2}$. Here the majorization relation $'\succ'$ means $\alpha_{0}^{2}\ge \beta_{0}^{2}$ and $\alpha_{0}^{2}+\alpha_{1}^{2}=\beta_{0}^{2}+\beta_{1}^{2}=1$. It is easy to check that the optimal transformation $\phi\to\psi$ between any pair of entangled states $\phi\succ\psi$ can be achieved by LOCC if the local UQF is available. And if a two-qubit unitary gate plays the local filtering role in the optimal conversion of all pair of entangled states, it is a UQF. Next our main goal is to find the necessary and sufficient conditions a UQF should satisfies.

{\bf Theorem}: Two-qubit unitary gate $U_{EA}$ is a UQF if and only if 
\bse
\bea
U_{EA}\circ U_{E}\otimes U_{A}|00\rangle&=&a_{0}|e_{0}\rangle^{E}|\eta\rangle^{A}+a_{1}|e_{1}\rangle^{E}|\mu\rangle^{A},\label{c1}\\
U_{EA}\circ U_{E}\otimes U_{A}|01\rangle&=&|e_{0}\rangle^{E}|\eta_{\bot}\rangle^{A},\label{c2}
\eea
\ese
hold for any value of $0\leq a_{0}\leq 1$ by tuning proper product unitary operator $U_{E}\otimes U_{A}$.

{\it Proof}:
\bea
\label{EAB}
|\phi\rangle^{EAB}&=&U_{EA}\circ U_{E}\otimes U_{A}|0\rangle^{E}|\phi\rangle^{AB} \nonumber\\
&=&\alpha_{0}|\phi_{0}\rangle^{EA}|0\rangle^{B}+\alpha_{1}|\phi_{1}\rangle^{EA}|1\rangle^{B}\nonumber\\
&=&\alpha_{0}(a_{0}|e_{0}\rangle^{E}|\phi_{0}\rangle^{A}+a_{1}|e_{1}\rangle^{E}|\phi_{1}\rangle^{A})|0\rangle^{B} +\alpha_{1}(b_{0}|e_{0}\rangle^{E}|\psi_{0}\rangle^{A}+b_{1}|e_{1}\rangle^{E}|\psi_{1}\rangle^{A})|1\rangle^{B} \nonumber\\
&=&|e_{0}\rangle^{E}(\alpha_{0}a_{0}|\phi_{0}\rangle^{A}|0\rangle^{B}+\alpha_{1}b_{0}|\psi_{0}\rangle^{A}|1\rangle^{B}) +|e_{1}\rangle^{E}(\alpha_{0}a_{1}|\phi_{1}\rangle^{A}|0\rangle^{B}+\alpha_{1}b_{1}|\psi_{1}\rangle^{A}|1\rangle^{B})
\eea
where $\langle\phi_{0}|\phi_{1}\rangle^{EA}=0$ and $\{|e_{0}\rangle,|e_{1}\rangle\}$ is the measurement on the ancilla. Suppose the transformation is successful if the measurement outcome is "0" and fails if the outcome is "1". Since the probability of success is optimal, we conclude that the failing state $\alpha_{0}a_{1}|\phi_{1}\rangle^{A}|0\rangle^{B}+\alpha_{1}b_{1}|\psi_{1}\rangle^{A}|1\rangle^{B}$ must be a product state, or else it can be converted to the goal state with some probability that contradicts the optimal transformation. In addition, we know that the conversion can be completed by only Alice'e operation. So we obtain
\bse
\bea
\alpha_{0}a_{1}|\phi_{1}\rangle^{A}|0\rangle^{B}+\alpha_{1}b_{1}|\psi_{1}\rangle^{A}|1\rangle^{B} &=&\sqrt{p_{1}}|\mu\rangle^{A}|\nu\rangle^{B},\label{fail}\\
\alpha_{0}a_{0}|\phi_{0}\rangle^{A}|0\rangle^{B}+\alpha_{1}b_{0}|\psi_{0}\rangle^{A}|1\rangle^{B} &=&\sqrt{p_{0}}(\beta_{0}V_{A}|0\rangle^{A}|0\rangle^{B}+\beta_{1}V_{A}|1\rangle^{A}|1\rangle^{B})\label{succ}
\eea
\ese
From {\it Eq.}(\ref{EAB}), we know that 
\be
\rho_{B}=Tr_{EA}(|\phi\rangle\langle\phi|)^{EAB}=p_{0}(\beta_{0}^{2}|0\rangle\langle0| +\beta_{1}^{2}|1\rangle\langle1|)+p_{1}|\nu\rangle\langle\nu|.
\ee
Combining with the condition $\rho_{B}=\alpha_{0}^{2}|0\rangle\langle0| +\alpha_{1}^{2}|1\rangle\langle1|$, we get $|\nu\rangle=|0\rangle$. From {\it Eqs.}(\ref{fail},\ref{succ}), we obtain $b_{1}=0,b_{0}=1$, $|\phi_{1}\rangle^{A}=|\mu\rangle^{A}$, and $\langle\phi_{0}|\psi_{0}\rangle^{A}=0$ that makes $\langle\phi_{0}|\phi_{1}\rangle^{EA}=0$ satisfied. Let us denote $|\phi_{0}\rangle^{A}=|\eta\rangle^{A}$ and $|\psi_{0}\rangle^{A}=|\eta_{\bot}\rangle^{A}$. Now {\it Eq.}(\ref{EAB}) is of the form
\bea
U_{EA}\circ U_{E}\otimes U_{A}|0\rangle^{E}|\phi\rangle^{AB} &=&\alpha_{0}(a_{0}|e_{0}\rangle^{E}|\eta\rangle^{A}+a_{1}|e_{1}\rangle^{E}|\mu\rangle^{A})|0\rangle^{B} +\alpha_{1}|e_{0}\rangle^{E}|\eta_{\bot}\rangle^{A}|1\rangle^{B} \nonumber\\
&=&|e_{0}\rangle^{E}(\alpha_{0}a_{0}|\eta\rangle^{A}|0\rangle^{B}+\alpha_{1}|\eta_{\bot}\rangle^{A}|1\rangle^{B}) +\alpha_{0}a_{1}|e_{1}\rangle^{E}|\mu\rangle^{A}|0\rangle^{B}
\label{uqf}
\eea
The proof is completed. Two immediate corollaries are the following.

{\bf Corollary 1} The entanglement capacity of the interaction $U_{EA}$ of a UQF is at least one ebit. 

Notice that when $a_{0}\to 0$, 
\be
U_{EA}\circ U_{E}\otimes U_{A}|0\rangle^{E}(|00\rangle^{AB}+|11\rangle^{AB})\to |e_{1}\rangle^{E}|\mu\rangle^{A}|0\rangle^{B}+|e_{0}\rangle^{E}|\eta_{\bot}\rangle^{A}|1\rangle^{B}.
\ee
The entanglement between $E$ and $AB$ is one ebit. So the entanglement capacity of $U_{EA}$ is at least one ebit.

{\bf Corollary 2} If $U_{EA}$ is an interaction unitary gate of UQF, then $U_{E}$ and $U_{A}$ can be found such that
\bse
\bea
U_{EA}\circ U_{E}\otimes U_{A}|00\rangle&=&|e_{0}\rangle^{E}|\eta\rangle^{A},\\
U_{EA}\circ U_{E}\otimes U_{A}|01\rangle&=&|e_{0}\rangle^{E}|\eta_{\bot}\rangle^{A}.
\eea
\ese
This conclusion comes directly from $a_{0}=1$.

All the above discussion is based on the assumption that there exists UQF unitary gate. Does such interaction gate really exist and how do we find such a gate? The answer to the existence problem is YES and a simple choice is the CNOT gate. Before we prove CNOT is an eligible candidate, we show the general scheme to search the UQF gate.

Any two-qubit operator can be decomposed as its canonical form \cite{Kraus}
\bea
U_{EA}&=&V_{E}\otimes V_{A}U_{d}U_{E}\otimes U_{A},\nonumber\\
U_{d}&=&\exp{i(\sum_{k=1}^{3}\alpha_{k}\sigma_{k}^{E}\otimes \sigma_{k}^{A})},
\eea
where $\sigma_{k}$ are the Pauli operators and $\pi/4\ge\alpha_{1}\ge\alpha_{2}\ge|\alpha_{3}|$. $U_{d}$ is the interaction term. Because arbitrary product unitary operator is allowed before and after the implementation of $U_{EA}$, it is $U_{d}$ that plays the role of UQF. So we can just consider $U_{EA}$ as the form $U_{d}(\alpha_{k})$. A generic $SU(2)$ operator is dependent on three parameters. The global phase can be ignored since it has no effect in the transformation. Denote $U_{E}(\alpha_{4},\alpha_{5})$ and $U_{A}(\alpha_{6},\alpha_{7})$. 
\bse
\bea
U_{d}\circ U_{E}\otimes U_{A}|00\rangle=|\Phi(\alpha_{i})\rangle^{EA},\\
U_{d}\circ U_{E}\otimes U_{A}|01\rangle=|\Psi(\alpha_{i})\rangle^{EA}.
\eea
\ese

From the theorem, $|\Psi(\alpha_{i})\rangle^{EA}=|e_{0}\rangle^{E}|\eta_{\bot}\rangle^{A}$ is a product state that gives a equation satisfied by $\alpha_{i}$, which we denote the constraint as $C1$. Then,
\bea
|e_{0}\rangle^{E}\langle e_{0}|=Tr_{A}|\Psi(\alpha_{i})\rangle\langle\Psi(\alpha_{i})|,\\
|a_{0}|^{2}=Tr_{A}\langle e_{0}|\Phi(\alpha_{i})\rangle\langle\Phi(\alpha_{i})|e_{0}\rangle =Tr_{EA}[(|\Phi(\alpha_{i})\rangle\langle\Phi(\alpha_{i})|) Tr_{A}(|\Psi(\alpha_{i})\rangle\langle\Psi(\alpha_{i})|)]
\eea
So $a_{0}=a_{0}(\alpha_{i})$. The problem of searching UQF is reduced to find the range of $a_{0}$ under the constraint $C1$. Fixing three interaction parameters, if the range of $|a_{o}|$ is $[0,1]$, $U_{d}$ is an eligible UQF.

It is a tedious job to search all the candidates of UQF by the general scheme. Now even it does not give a definite answer to the existence problem, though UQF should exist since we have more parameters to be tuned. Note that our main aim is to find a simple UQF that can be easily implemented in experiment. Even if all the UQFs are found, the simplest one will be chosen. The two corollaries drop a hint that probably CNOT is an eligible one since CNOT is the simplest interaction unitary operator that satisfies corollary 1 and corollary 2. Also it does not contradict with {\it Eq.}(\ref{c1}), since $|\mu\rangle^{A}$ is unnecessarily orthogonal to $|\eta\rangle^{A}$. Hopefully. we might as well have a try with the CNOT gate. Luckily enough, it really works. Now we show how it works. Choose $U_{A}=\sigma_{1}$ and $U_{E}=(\cos\theta|0\rangle+\sin\theta|1\rangle)\langle0|+(\sin\theta|0\rangle-\cos\theta|1\rangle)\langle1| $ where the parameter $\theta$ is determined by the parameter of a specific filter $F(\alpha)$, then
\bse
\bea
U_{CNOT}(\cos\theta|0\rangle+\sin\theta|1\rangle)(|0\rangle+|1\rangle)/\sqrt{2} &=&(\cos\theta|0\rangle+\sin\theta|1\rangle)(|0\rangle+|1\rangle)/\sqrt{2}\nonumber\\
&=&\cos2\theta|e_{0}\rangle|+\rangle+\sin2\theta|e_{1}\rangle|+\rangle,\\
U_{CNOT}(\cos\theta|0\rangle+\sin\theta|1\rangle)(|0\rangle-|1\rangle)/\sqrt{2} &=&(\cos\theta|0\rangle-\sin\theta|1\rangle)(|0\rangle-|1\rangle)/\sqrt{2}\nonumber\\
&=&|e_{0}\rangle|-\rangle,
\eea
\ese
with
\bse
\bea
|e_{0}\rangle&=&\cos\theta|0\rangle-\sin\theta|1\rangle,\label{e0}\\
|e_{1}\rangle&=&\sin\theta|0\rangle+\cos\theta|1\rangle,\label{e1}\\
|+\rangle&=&1/\sqrt{2}(|0\rangle+|1\rangle),\\
|-\rangle&=&1/\sqrt{2}(|0\rangle-|1\rangle).
\eea
\ese
Then {\it Eq.}(\ref{uqf}) gives
\be
U^{CNOT}_{EA}\circ U_{E}\otimes U_{A}(\alpha_{0}|00\rangle^{AB}+\alpha_{1}|11\rangle^{AB}) =|e_{0}\rangle^{E}(\alpha_{0}\cos2\theta|+\rangle^{A}|0\rangle^{B}+\alpha_{1}|-\rangle^{A}|1\rangle^{B}) +\alpha_{0}\sin2\theta|e_{1}\rangle^{E}|+\rangle^{A}|0\rangle^{B}
\ee
So $\{U_{EA},U_{E},U_{A},M_{E}\}$ plays the role a a universal quantum filter, where $M_{E}$ is the measurement in the basis $\{|e_{0}\rangle,|e_{1}\rangle\}$. Also we find another UQF with $U_{d}=\pi/4\sigma_{1}\otimes\sigma_{1}+\pi/8\sigma_{2}\otimes\sigma_{2}+\pi/8\sigma_{3}\otimes\sigma_{3}$ and proper product unitary transformation $V_{E}\otimes V_{A}$ and measurement $N_{E}$, albeit their parameters are complicatedly dependent on each other and the interaction parameters $\alpha_{k}$. A reasonable conjecture is that any interaction $U_{d}$ with one ebit entanglement capacity is an eligible candidate of UQF.

As an example, we use the CNOT UQF to optimally convert an entangled state $|\phi\rangle=\cos x|00\rangle^{AB}+\sin x|11\rangle^{AB}$ to $|\psi\rangle=\cos y |00\rangle^{AB}+\sin y|11\rangle^{AB}$ where $0\leq x\leq y\leq \pi/4$ that satisfies $\phi\succ\psi$. First, prepare the ancilla state $\cos\theta|0\rangle+\sin\theta|1\rangle$ with $\theta$ fixed by $\cos2\theta=\tan x/\tan y$. Rotate the basis $\{|0\rangle^{A},|1\rangle^{A}\}$ to $\{|+\rangle^{A},|-\rangle^{A}\}$ by $\sigma_{1}^{A}$. Second, implement CNOT gate on the ancilla $E$ and system $A$. Third, measure the ancilla in the basis $\{|e_{0}\rangle,|e_{1}\rangle\}$ of {\it Eq.}(\ref{e0},\ref{e1}). It is easy to check that $p(\phi\to\psi)=\sin^{2} x/\sin^{2} y$. So the optimal transformation is achieved. If the goal state is a MES, any non-MES is transformed to a MES with maximal probability by the CNOT UQF.

\section{Universal quantum diluter}
Another exciting fact is that the CNOT filter can be employed as a universal quantum diluter (UQD). Similarly, a UQD is defined as a device that can dilute $(\alpha_{0},\alpha_{1})$ to $(\beta_{0},\beta_{1})$ with certainty when $\phi\prec\psi$ \cite{Nielsen}. The essential idea is that proper $U_{E}\otimes U_{A}$ and the measurement basis $M_{E}$ are chosen. Here we adopt Nielsen's dilution protocol \cite{Nielsen}. The first step of Nielsen's protocol is to transform $|\phi\rangle^{AB}$ to the form $|\phi'\rangle^{AB}=1/\sqrt{2}(|00\rangle+|1\rangle(\cos\gamma|0\rangle+\sin\gamma|1\rangle))$ by unitary action on system $A$ where $\gamma$ is chosen to satisfy $\alpha_{0}^{2}=(1+\cos\gamma)/2$. The second step is to perform the POVM described by $\{M_{0}, M_{1}\}$ in the basis $\{|0\rangle^{A},|1\rangle^{A}\}$,
\be
M_{0}=\left[\begin{array}{cc}\cos\delta & 0\\0 &\sin\delta\end{array}\right],~~~ M_{1}=\left[\begin{array}{cc}\sin\delta & 0\\0 &\cos\delta\end{array}\right]
\ee  
where $\delta$ is determined as $\delta=1/2\arcsin{(2(\beta_{0}^{2}-\beta_{0}^{4})^{1/2}/\sin\gamma)}$. Then the resulting state after the measurement is equivalent to the goal state by local unitary operations. As pointed out in \cite{Nielsen}, the POVM $\{M_{0},M_{1}\}$ can be implemented using projective measurement and bipartite unitary transformation. Here we demonstrate that any POVM of the form $\{M_{0},M_{1}\}$ could be realized by the CNOT transformation supplemented with $SU(2)$ and projective measurement. Now we show how the UQD works. First, the same transformation $U_{EA}\circ U_{E}\otimes U_{A}$ as UQF is performed on $|0\rangle^{E}|\phi'\rangle^{AB}$ which gives
\be
U_{EA}^{CNOT}\circ U_{E}\otimes U_{A}|0\rangle^{E}|\phi'\rangle^{AB} =(\cos\theta|0\rangle^{E}+\sin\theta|1\rangle^{E})|+\rangle^{A}|0\rangle^{B} +(\cos\theta|0\rangle^{E}-\sin\theta|1\rangle^{E})|-\rangle^{A}(\cos\gamma|0\rangle^{B}+\sin\gamma|1\rangle^{B})
\ee
If $\theta$ is chosen to satisfy $\cos2\theta=\sin2\delta$, then there exists unitary transformation $V_{E}$ such that
\bse
\bea
V_{E}(\cos\theta|0\rangle^{E}+\sin\theta|1\rangle^{E})=\cos\delta|0\rangle^{E}+\sin\delta|1\rangle^{E},\\
V_{E}(\cos\theta|0\rangle^{E}-\sin\theta|1\rangle^{E})=\sin\delta|0\rangle^{E}+\cos\delta|1\rangle^{E}.
\eea
\ese
Next, Alice applies $V_{E}\otimes U_{A}$ to the ancilla and system $A$, then
\bea
&~&V_{E}\otimes U_{A}\circ U_{EA}^{CNOT}\circ U_{E}\otimes U_{A}|0\rangle^{E}|\phi'\rangle^{AB}\nonumber\\ &=&(\cos\delta|0\rangle^{E}+\sin\delta|1\rangle^{E})|0\rangle^{A}|0\rangle^{B} +(\sin\delta|0\rangle^{E}+\cos\delta|1\rangle^{E})|1\rangle^{A}(\cos\gamma|0\rangle^{B} +\sin\gamma|1\rangle^{B})\nonumber\\
&=&|0\rangle^{E}(\cos\delta|0\rangle^{A}|0\rangle^{B}+\sin\delta|1\rangle^{A}(\cos\gamma|0\rangle^{B} +\sin\gamma|1\rangle^{B}))\nonumber\\ &+&|1\rangle^{E}(\sin\delta|0\rangle^{A}|0\rangle^{B}+\cos\delta|1\rangle^{A}(\cos\gamma|0\rangle^{B} +\sin\gamma|1\rangle^{B}))
\eea
Finally, Alice measures the ancilla in the basis $\{|0\rangle^{E},|1\rangle^{E}\}$ and the POVM is realized. So the CNOT diluter is universal. Most known implementations of non-local operations either require MES or become probabilistic with non-MES. Nevertheless sometimes non-MES is required in some quantum tasks. An example related with our topic is the remote POVMs discussed in \cite{Reznik}.

\section{general two-outcome POVM}
From the above discussion, we know that CNOT gate can be utilized to perform two-outcome POVMs of UQF and UQD with the aid of qubit unitary operator and projective measurement. These two types belong to a larger subset of POVM $\{D_{0}, D_{1}\}$ of the form
\be
\label{povm}
D_{0}=\left[\begin{array}{cc}\cos\alpha & 0\\0 &\cos\beta\end{array}\right],~~~ D_{1}=\left[\begin{array}{cc}\sin\alpha & 0\\0 &\sin\beta\end{array}\right]
\ee
A general two-outcome POVM is described by $\{M_{0}, M_{1}\}$ where $M_{i}$ is a generic $2\times 2$ matrix satisfying the completeness relation $M_{0}^{\dagger}M_{0}+M_{1}^{\dagger}M_{1}=I$. It is unitarily equivalent to $\{N_{0}, N_{1}\}$ of the form
\be
N_{0}=V_{0}\left[\begin{array}{cc}\cos\alpha & 0\\0 &\cos\beta\end{array}\right],~~~ N_{1}=V_{1}\left[\begin{array}{cc}\sin\alpha & 0\\0 &\sin\beta\end{array}\right],
\ee 
where $V_{0},V_{1}$ are two unitary operators. The proof is simple. From matrix analysis and the completeness relation, we know that $M_{0}^{\dagger}M_{0}$ and $M_{1}^{\dagger}M_{1}$ can be diagonalized by the same unitary operator $U$ and $M_{i}=V_{i}D_{i}U$. Reducing $U$, $\{N_{0}, N_{1}\}$ is obtained.

Actually, a general two-outcome POVM could be realized by CNOT operator. 
\be
U_{EA}^{CNOT}\circ U_{E}\otimes U_{A}|0\rangle^{E}|\phi\rangle^{AB} =\alpha_{0}(\cos\theta|0\rangle^{E}+\sin\theta|1\rangle^{E})|+\rangle^{A}|0\rangle^{B} +\alpha_{1}(\cos\theta|0\rangle^{E}-\sin\theta|1\rangle^{E})|-\rangle^{A}|1\rangle^{B},
\ee
where $\theta$ is fixed as $\cos2\theta=\cos(\alpha-\beta)$. There exists unitary operator $V_{E}$ such that
\bse
\bea
V_{E}(\cos\theta|0\rangle^{E}+\sin\theta|1\rangle^{E})=\cos\alpha|0\rangle^{E}+\sin\alpha|1\rangle^{E},\\
V_{E}(\cos\theta|0\rangle^{E}-\sin\theta|1\rangle^{E})=\cos\beta|0\rangle^{E}+\sin\beta|1\rangle^{E}.
\eea
\ese
Under the operator $V_{E}\otimes U_{A}$,
\bea
V_{E}\otimes U_{A}\circ U_{EA}^{CNOT}\circ U_{E}\otimes U_{A}|0\rangle^{E}|\phi\rangle^{AB} &=&|0\rangle^{E}(\alpha_{0}\cos\alpha|0\rangle^{A}|0\rangle^{B}+\alpha_{1}\cos\beta|1\rangle^{A}|1\rangle^{B})
\nonumber\\
&+&|1\rangle^{E}(\alpha_{0}\sin\alpha|0\rangle^{A}|0\rangle^{B}+\alpha_{1}\sin\beta|1\rangle^{A}|1\rangle^{B})
\eea
Next, measurement is performed on the ancilla in the basis $\{|0\rangle^{E},|1\rangle^{E}\}$. Finally, Alice performs $V_{i}$ on the system $A$ when $i$ outcome is obtained. So the general POVM $\{N_{0}, N_{1}\}$ is realized. As a matter of fact, it is {\it Eq.}(\ref{povm}) that plays the role in entanglement conversion.

\section{optical realization}
The spin of the photon is utilized as a qubit in quantum communication. There have been extensive work on the two entangled photons. The creation of two-photon entangled state is easily available. It is hard to handle one of the two entangled photons with another new photon that acts as the ancilla qubit. The realization of CNOT operator between the spins of two photons is a challenging task. However, CNOT operations can be easily implemented between two different degree of freedom of single photon \cite{Chen}. We can choose the path modes of one photon as the ancilla system. So optical UQF could be realized experimentally with current technology.\\

\section{summary}
In summary, we introduce universal quantum filter (diluter) that can realize optimal conversion between any two pure entangled states. The optical realization in experiment is shortly discussed. Maybe UQF will soon appear in the toolbox as a standard device since the filtering process is one basic transformation in quantum communication.



\begin{thebibliography}{99}

\bibitem{NC}
M. A. Nielsen and I. L. Chuang, {\it Quantum Computation and Quantum Information} (Cambridge University Press, Cambridge, 2000).

\bibitem{Bennett2}
C. H. Bennett, G. Brassard, S. Popescu, B. Schumacher, J. A. Smolin, W. K. Wootters
, Phys. Rev. Lett. {\bf 76}, 722 (1996); C. H. Bennett, D. P. DiVincenzo, J. A. Smolin and W. K. Wootters, Phys. Rev. A {\bf 54}, 3824 (1996).

\bibitem{Popescu}
S. Popescu, Phys. Rev. Lett. {\bf 74}, 2619 (1995).

\bibitem{Gisin}
N. Gisin, Phys. Lett. A  {\bf 210}, 151 (1996).

\bibitem{Horodecki}
M. Horodecki, P. Horodecki, and R. Horodecki, Phys. Rev. Lett. {\bf 78}, 574 (1997).

\bibitem{Vaetraete}
F. Verstraete, J. Dehaene, B. D. Moor, Phys. Rev. A {\bf 64}, 010101(R) (2001); F. Verstraete, H. Verschelde, Phys. Rev. Lett. 90, 097901 (2003).

\bibitem{Nielsen}
M. A. Nielsen, Phys. Rev. Lett. {\bf 83}, 436 (1999).

\bibitem{Vidal}
G. Vidal, Phys. Rev. Lett. {\bf 83}, 1046 (1999).

\bibitem{Kraus}
B. Kraus and J. I. Cirac, Phys. Rev. A {\bf 63}, 062309 (2001).

\bibitem{Reznik}
B. Reznik, {\it Remote generalized measurements (POVMs) require non-maximal entanglement}, quant-ph/0203055.

\bibitem{Chen}
Z.-B. Chen, J.-W. Pan, Y.-D. Zhang, C. Brukner, A. Zeilinger, Phys. Rev. Lett. {\bf 90}, 160408 (2003).

\end{thebibliography}
\end{document}